\documentclass[sigconf]{acmart}
\AtBeginDocument{%
  }

\copyrightyear{2025}
\acmYear{2025}
\setcopyright{cc}
\setcctype{by}
\acmConference[SC Workshops '25]{Workshops of the International Conference for High Performance Computing, Networking, Storage and Analysis}{November 16--21, 2025}{St Louis, MO, USA}
\acmBooktitle{Workshops of the International Conference for High Performance Computing, Networking, Storage and Analysis (SC Workshops '25), November 16--21, 2025, St Louis, MO, USA}
\acmDOI{10.1145/3731599.3767554}
\acmISBN{979-8-4007-1871-7/2025/11}

\usepackage{hyperref}
\usepackage{listings}
\usepackage{xcolor}
\usepackage{framed,xcolor}
\makeatletter
\@ifundefined{FrameSep}{\newlength{\FrameSep}}{} 
\makeatother
\colorlet{hpcback}{black!10}        
\newenvironment{hpcbox}{%
  \MakeFramed{\advance\hsize-\width \FrameRestore}%
}{\endMakeFramed}

\usepackage[htt]{hyphenat}

\lstdefinelanguage{json}{
    morestring=[b]",
    morestring=[d]',
    stringstyle=\color{teal},
    comment=[l]{//},
    morecomment=[s]{/*}{*/},
    commentstyle=\color{gray},
    morekeywords={
        true,false,null
    },
    keywordstyle=\color{blue},
    showstringspaces=false,
    literate=
     *{:}{{\textcolor{orange}{:}}}1
      {,}{{\textcolor{orange}{,}}}1
      {[}{{\textcolor{orange}{[}}}1
      {]}{{\textcolor{orange}{]}}}1
      {\{}{{\textcolor{orange}{\{}}}1
      {\}}{{\textcolor{orange}{\}}}}1
}

\begin{document}

\title{An HPC-Inspired Blueprint for a Technology-Agnostic \\Quantum Middle Layer}

\author{Stefano Markidis, Gilbert Netzer, Luca Pennati, Ivy Peng}
\affiliation{%
  \institution{KTH Royal Institute of Technology}
  \city{Stockholm}
  \country{Sweden}
}

\renewcommand{\shortauthors}{Markidis et al.}

\begin{abstract} We present a blueprint for a quantum middle layer that supports applications across various quantum technologies. Inspired by concepts and abstractions from HPC libraries and middleware, our design is backend-neutral and context-aware. A program only needs to specify its intent once as typed data and operator descriptors. It declares what the quantum registers mean and which logical transformations are required, without committing to gates, pulses, continuous-variable routines, or anneal backend. Such execution details are carried separately in a context descriptor and can change per backend without modifying the intent artifacts.

We develop a proof of concept implementation that uses JSON files for the descriptors and two backends: a gate-model path realized with IBM Qiskit Aer simulator and an annealing path realized with D-Wave Ocean’s simulated annealer. On a Max-Cut problem instance, the same typed problem runs on both backends by varying only the operator formulation (Quantum Approximated Optimization Algorithm formulation vs. Ising Hamiltonian formulation) and the context. The proposed middle layer concepts are characterized by portability, composability, and its minimal core can evolve with hardware capabilities.

\end{abstract}




\maketitle

\section{Introduction}
The broader availability of quantum computing systems has led to the development of diverse programming approaches for expressing quantum algorithms, applications, and workflows. To support emerging quantum computing platforms, several programming frameworks have rapidly emerged, mainly driven by vendors targeting specific quantum computing systems.
For qubit-based systems using quantum circuits IBM's \texttt{Qiskit}~\cite{mckay2018qiskit} and
Google's \texttt{Cirq}~\cite{isakov2021simulations} have become
the de facto approach.
Other frameworks target pulse-level programming for analog quantum computers (Pasqal's \texttt{Pulser}~\cite{silverio2022pulser}), differentiable quantum computing (Xanadu's \texttt{PennyLane}~\cite{bergholm2018pennylane}) or continuous variable quantum computing (Xa\-na\-du's \texttt{Strawberry Fields}~\cite{killoran2019strawberry}).
Quantum annealers and adiabatic computing have also emerged as an essential and viable approach for solving optimization problems (for instance, D-Wave \texttt{Ocean}~\cite{hassan2019c}).
Advances in algorithms that use measurement-in-the-middle and dynamic feedback have led to new developments in existing quantum assembly languages, 
such as \texttt{QASM 3.0}~\cite{cross2022openqasm}, and to novel
approaches, such as QHDL~\cite{netzer2026a}, with emphasis on tight
integration between classical and quantum computations.
This rapid proliferation has led to a diverse, creative, yet unstructured approach to design programming interfaces for quantum computers.

In this work, we want to tackle a critical and challenging software
engineering problem:
\emph{How do we design a layered quantum software stack whose components are agnostic to the underlying quantum computer and that is composable with other quantum software and classical systems which is minimal yet sufficient and extendable to describe quantum applications while offering performance portability?} 
We focus on the so-called middle layer, the software layer used by quantum applications to interface to compilers and transpilers.
The middle layer provides common program representations, device abstractions, libraries and runtimes to make programs portable by preparing workloads before the compiler/transpiler takes over.

To tackle this research challenge, we examine libraries and data models in the HPC field, which had similar challenges to quantum computing, e.g., supporting different heterogeneous computing technologies in a minimal and composable way. Historically, the availability of HPC systems to researchers increased significantly in the mid-1980s, leading to several approaches to program supercomputers. As a result of intense experimentation, strategies, and standardization efforts, a few critical building blocks have emerged. An example is the Message Passing Interface (MPI)~\cite{gropp1999using}, which replaced all other initial efforts to develop communication libraries for supercomputers. Other well-established HPC libraries include \texttt{BLAS} and \texttt{LAPACK} for linear algebra~\cite{blackford2002updated}, \texttt{PETSc} for linear and nonlinear solvers~\cite{mills2021toward}, \texttt{FFTW} for fast Fourier transforms~\cite{frigo1998fftw}, and \texttt{ADIOS} for I/O~\cite{godoy2020adios}. We apply the lessons learned from HPC in 
developing portable, technology-agnostic software to the design of the quantum middle layer.

We summarize our contributions in this work as follows:
\begin{itemize}
\item We design a back-end-neutral, context-aware quantum middle layer, inspired by HPC middleware. The quantum middle layer blueprint specifies quantum data type descriptors for explicit register meaning, quantum operator descriptors with parameters, result schemas, and optional device-independent cost hints, and orthogonal context for execution policy. 
\item We implement a proof of concept using JSON as the interchange format and two backends: a gate-model path (IBM Qiskit Aer) that produces a Quantum Approximated Optimization Algorithm (QAOA) descriptor sequence, and a quantum annealing path (D-Wave Ocean) that produce an Ising/Binary Quadratic Model (BQM) operator descriptor. \item We demonstrate the proposed solution by solving the Max-Cut instance with a shared quantum data type and explicit decoding schema, showing portability across different backends.
\end{itemize}

\section{A Motivational Example}
To ground our discussion in a practical, well-defined use case, we compare the
computation of the Discrete Fourier Transform (DFT) 
with performing it's quantum counterpart, 
the Quantum Fourier Transform (QFT)~\cite{nielsen2001quantum}.

A common approach in HPC is to use the \texttt{FFTW} library~\cite{frigo1998fftw} and its data types and abstractions.
To do so, the application describes the desired data type (real, complex), array layout (strides, sizes) and transformation parameters (direction, number of dimensions, length per dimension) to the library that
creates an optimal \emph{plan}, which can later be executed by a suitable library call to compute the
desired transformations.


\begin{lstlisting}[float=tp,
                   language=python,
                   basicstyle=\ttfamily\footnotesize,
                   numbers=none,
                   caption={Qiskit Code for a 10-qubit QFT},
                   label={lst:qskit}]
#imports
...
# 1. Middle Layer: a 10-qubit QFT + Measure  
qc = QuantumCircuit(10, 10)
qft = QFT(num_qubits=10, 
          approximation_degree=0, 
          do_swaps=True)
qc.append(qft.to_instruction(), range(10))
qc.measure(range(10), range(10))

# 2. Backend: Compile and run on a simulator
backend = Aer.get_backend("aer_simulator")
tqc=transpile(qc, backend=backend, optimization_level=2)
result = backend.run(tqc, shots=10000).result()
print(result.get_counts())
\end{lstlisting}

In contrast, Listing~\ref{lst:qskit} shows a Python program to utilize the built-in QFT library from IBM's Qiskit framework to perform a QFT.
The first part of the program defines a quantum circuit to perform the QFT and is akin to a middle layer.
The constructor \texttt{qc = QuantumCircuit(10, 10)} creates an empty circuit with 10 qubits and 10 classical bits.
Independently, \texttt{qft = QFT(..., ..., ...)} creates a 10-qubit wide QFT operation with two options:
\texttt{approximation\_degree=0} requests an exact transformation, and \texttt{do\_swaps=True} requests
that the order of output qubits should be reversed similar to typical FFT algorithm~\cite{markidis2024quantum}.
\texttt{qc.append(..., range(10))} adds the QFT operation to the end of the circuit, with
\texttt{range(10)} mapping the circuits qubits (numbered 0, 1, ..., 9) to the QFT's 10 inputs and outputs
in the same order.
\texttt{qc.measure(...)} adds 10 single-qubit measurements in computational basis to the circuit, each storing the result in the matching classical bit (again numbered 0 to 9).
The second part of the program requests that the circuit be transpiled
into an optimized circuit (\texttt{tqc = transpile(...)}) suitable for the
requested state vector simulator (\texttt{Aer}) as specified by the \texttt{backend = ...} object.
This circuit is executed 1000 times, \texttt{result = backend.run(...).result()},
and the measurement results are displayed, \texttt{print(...)}.
%
%

In the context of a technology-agnostic middle layer we would like to point to a number of
potential issues with the presented example which our design intends to address:

\emph{The quantum program does not have typed data.} 
The user only includes the number of qubits, but there is no declaration of what the register means. For instance, there is no information about how the data is encoded in the quantum register. It is not defined whether the quantum states are interpreted as integer, signed integer, fixed point real number, Boolean, or phase encoding, to include a few potential interpretation of the information encoded in the quantum states. Moreover, the Endianess convention for mapping to quantum states is not explicitly expressed, causing the composability with other libraries to be problematic. Furthermore, the usage of \texttt{num\_qubits} would make it impractical for expressing QFT on non-qubit-based systems, such as continuous-variable quantum computing systems, which uses qumodes instead, or use Hamiltonian approaches that use a parametrization of Hamiltonian. 

\emph{The QFT circuit is generated without knowledge of the backend}
by the \texttt{qft.to\_instruction()} method.
This can be problematic if the execution context requires the usage of distributed quantum computing and the insert of teleportation~\cite{cardama2026netqir}, or the user requires Quantum Error Correction (QEC)~\cite{wong2022introduction} acting on the QFT formulation.
Deferring circuit generation until the back-end parameters are known would allow choosing optimized
QFT circuits instead.

\emph{The basis for carrying out and interpreting the measurement results is implicitly specified.}
Qiskit defaults to measure in the computational (Z) basis, and defers interpretation of the returned
result counts to the user.

\emph{There is no mechanism to express the execution policy.} For instance, a user might require a logical execution (usage of logical qubits instead of direct usage of physical qubits) when running QEC, multi-QPU execution, providing hints for pulse transpiling, or annealer settings. A technology-agnostic middle layer requires an explicit way to set the execution context of the QFT.

\emph{The reordering of the output qubits of the QFT is not visible to subsequent operations.}
This reordering affects the encoding of the information and the user has to adjust the following
operations accordingly.
This can also mask opportunities for automated optimizations, e.g. by rearranging follow up inverse
QFTs.

\emph{The cost information is not visible.} The \texttt{Qiskit} library does not expose cost metadata such as circuit depth, two-qubit gate count, ancilla demand, communication volume, or expected duration.%
, just to list some important cost metadata, for instance. 
Without this information, a scheduler cannot choose an appropriate back-end and topology, or estimate queue and runtime.
A technology-agnostic middle layer should include a \texttt{cost\_hint} to each operator, analogous to FLOP counts and communication estimates used by HPC schedulers.

 
\section{Design Principles \& Requirements}
The goal of this work is to provide a minimal set of middle layer abstractions to implement a quantum technology-agnostic middle layer. To ensure our middle layer is agnostic, the same program should target different quantum devices beyond gate-based systems, including quantum annealers and continuous variable systems. The support for diverse quantum targets requires \textit{late binding} at the backend level. The middle layer should allow late parameter binding and adaptive control (mid-circuit measurement and feedback), while forbidding implicit measurements or resource grabs. In addition, the execution context, e.g., the quantum computing platform in use and the error correction usage, for instance, is explicit to avoid hidden side effects. \textit{Composability} is a second design principle we follow to ensure that independently defined components requiring the usage of either different or the same resources can interoperate. To ensure composability, data and operators of the middle layer should be explicit and not inferred or set by default. For instance, results need unambiguous decoding rules (e.g., bit or mode ordering, datatype interpretation). Finally, the design should be \textit{minimalistic}, keeping the core abstractions and libraries minimal.

\begin{hpcbox}
\textbf{HPC Point-of-View}:
Several HPC middle layers follow these design principles. For instance, BLAS follows the minimalism design principle: it provides a small, stable kernel with precise data/layout contracts. MPI achieves composability with other libraries and interoperability with other HPC libraries like PETSc, via explicit context such as communicators and datatypes. When coming to late binding, PETSc and Trilinos, separate mathematical meaning of the operators, e.g., providing matrix and preconditioners shells, defining dimensions and algebra, from realization, e.g. solver and preconditioner choices are late-bound.
\end{hpcbox}

\section{Quantum Middle Layer Concepts and Components}
\begin{figure}[t]
    \centering
    \includegraphics[width=\linewidth]{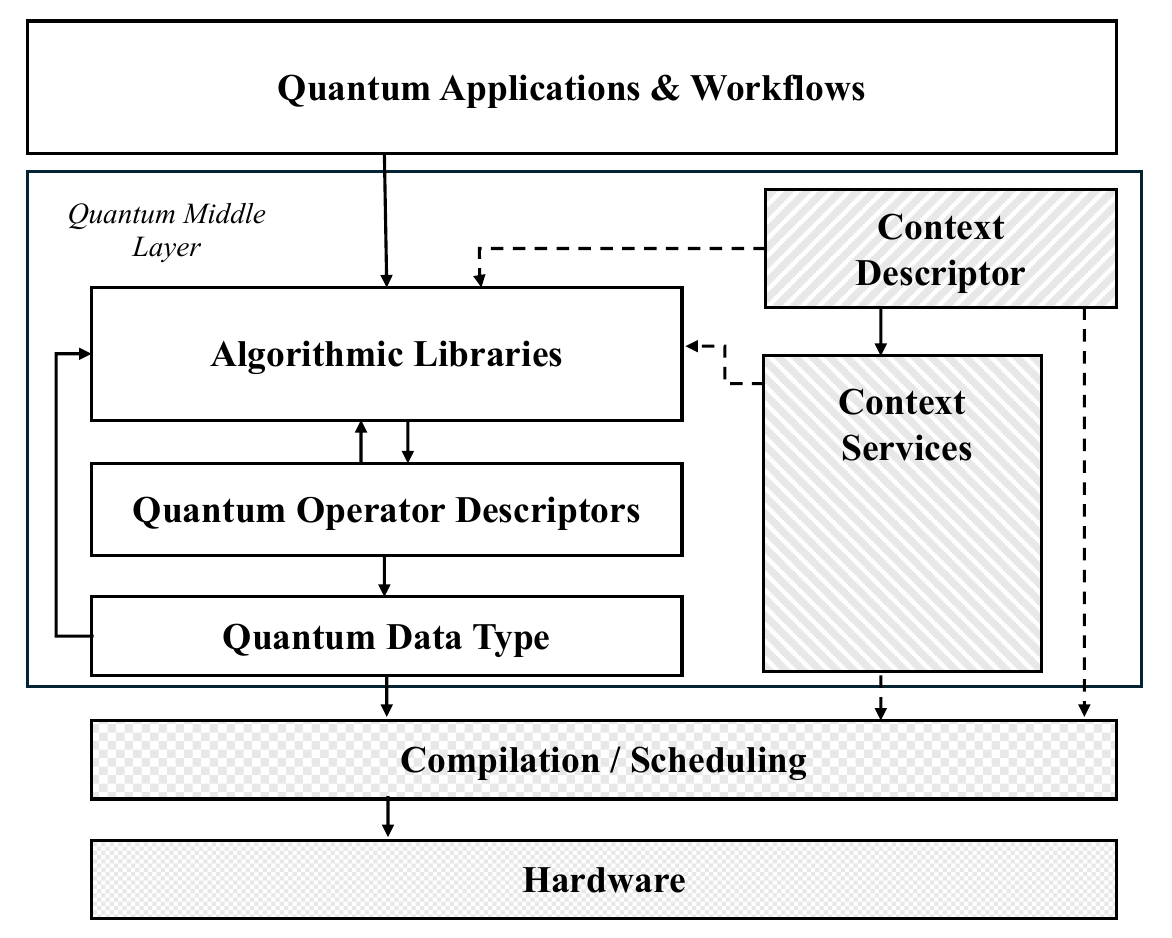}
    \caption{A diagram of a backend-neutral context-aware quantum middle layer.}
    \label{fig:middle_layer}
\end{figure}

This section presents the different components of the proposed backend-neutral context-aware quantum middle layer blueprint, as shown in Fig.~\ref{fig:middle_layer}. Our quantum middle layer has four main components: \emph{(i)} quantum data type descriptors to provide quantum registers with explicit meaning; \emph{(ii)} quantum operator descriptors to express logical transformations, independent of realization; \emph{(iii)} a context descriptor to describe the execution policy without changing semantics and context services; \emph{(iv)} algorithmic libraries to consume quantum data types and produce quantum operator descriptor sequences for algorithms. In the algorithmic libraries, a packaging step also bundles quantum data type, operators, and optional context for submission to a backend.
In the next subsections, we provide a more detailed description of the proposed quantum middle layer, using the example of QFT.

\subsection{Quantum Data Types}

A quantum data type is the semantic contract that tells every component what a quantum register means. Its objective is to allow independently written libraries to interpret registers identically -- no guessing about Endianess or number representation -- so that operators can compose safely, results can be decoded automatically, and validators can catch mismatches early. In addition, the quantum data type is hardware‑agnostic: it specifies what the data represents, but does not prescribe how to implement it, enabling the same program to be realized on different backends, including gates, pulses, CV, annealers, without changing upstream code.

The quantum register here can be of different kinds, depending on the interpretation of the amplitude of quantum states. Integer registers are a quantum data type interpreted with \texttt{AS\_INT} semantics and a declared bit significance (e.g., little vs big Endian), so the basis state $\lvert k\rangle$ decodes to the integer $k\in[0,2^n\!-\!1]$. Together with integers, other data types include Boolean registers, which encode $\{0,1\}$ with \texttt{AS\_BOOL} semantics for control logic and Quadratic Unconstrained Binary Optimization (QUBO)/Ising variables. Another example is \emph{phase register} that represents a fixed-point phase accumulator. In this quantum register, the index $k$ denotes a fraction of a complete turn according to a stated phase scale, with measurements interpreted as \texttt{AS\_PHASE}. 

Listing~\ref{lst:QFT_data_JSON} presents an example of a quantum data type in JSON format for a quantum data register for the QFT computation. The \texttt{\$schema} field names the JSON Schema used to validate the descriptor. The \texttt{id} and \texttt{name} identify the logical register. \texttt{width = 10} declares 10 logical carriers (typically qubits, but they could be qumodes in continuous-variable quantum computing settings), encoding fixed-point phase on the unit circle with resolution $1/1024$. The \texttt{encoding\_kind = PHASE\_REGISTER} indicates that the computational basis state \(|k\rangle\) denotes the discrete phase value associated with index $k$. \texttt{bit\_order = LSB\_0} fixes the significance order so index \(i\) has weight \(2^i\). \texttt{measurement\_semantics = AS\_PHASE} communicates to downstream tools to interpret Z-basis outcomes as a phase value. Finally, \texttt{phase\_scale = 1/1024} defines the mapping from the observed integer $k$ to a unitless phase fraction $k/1024$ of a full turn; multiply by $2\pi$ for radians or by $360$ for degrees. 

\begin{lstlisting}[float=tp,
                   language=json,
                   basicstyle=\ttfamily\footnotesize,
                   numbers=none,
                   caption={Quantum data type to be consumed by a QFT library},
                   label={lst:QFT_data_JSON}]
{
    "$schema": "qdt-core.schema.json",
    "id": "reg_phase",
    "name": "phase",
    "width": 10,
    "encoding_kind": "PHASE_REGISTER",
    "bit_order": "LSB_0",
    "measurement_semantics": "AS_PHASE",
    "phase_scale": "1/1024"
}
\end{lstlisting}
\begin{hpcbox}
\textbf{HPC Point-of-View}: Data types are the main characteristics of HPC libraries. For instance, MPI~\cite{gropp1999using} includes MPI Datatypes (incl. derived subarray/struct types). The MPI libraries make layout, element type, and ordering explicit so independent codes agree on interpretation. I/O libraries, such as HDF5~\cite{folk2011overview} and ADIOS2~\cite{godoy2020adios}, allow metadata to declare element type, shape, and attributes so that data can be written/read consistently across tools. PETSc is another example of leveraging data types, such as \texttt{Vec}, \texttt{IS}, and \texttt{DM}. These data types allow users to define the global index space, field layout, and distribution so that solvers interpret vectors/matrices the same way.
\end{hpcbox}

\subsection{Quantum Operator Descriptors}
A quantum operator is the mathematical rule that transforms one quantum state into another. The operator is the abstract action, independent of gates or their actual implementation. Examples of quantum operators are the QFT, a modular adder that is a primitive to add two qubits integers modulo a prime modulus, which is a main component of the Shor algorithm, or the Ising evolution operator that evolves a set of qubits under an Ising Hamiltonian for a specified time. In the middle layer, the operator's main goal is to instruct higher-level algorithms on the logical change that must occur (add two numbers, perform a QFT, or apply an Ising Hamiltonian), so that lower layers can determine how to implement it on the available hardware in the best way.

Listing~\ref{lst:Operator_JSON} shows descriptor for a Quantum Fourier Transform on the logical register \texttt{reg\_phase}. The field \texttt{rep\_kind = "QFT\_TEMPLATE"} identifies the logical transformation and that it is a realizable template rather than concrete gates. \texttt{domain\_qdt} (input register) and \texttt{codomain\_qdt} (output register) both reference \texttt{reg\_phase}, meaning the transform is in-place at the logical level. The \texttt{params} groups \texttt{approx\_degree = 0} asks for the exact QFT.  \texttt{do\_swaps = true} requests the final wire-reversal swaps. \texttt{inverse = false} selects the forward  QFT as opposed to the inverse. The \texttt{cost\_hint} provides an estimate (here, roughly 45 two-qubit gates and depth near 100), which backends may use for early planning and scheduling.

\begin{lstlisting}[float=tp,
                   language=json,
                   basicstyle=\ttfamily\footnotesize,
                   numbers=none,
                   caption={Operator Descriptor},
                   label={lst:Operator_JSON}]
{
    "$schema": "qod.schema.json",
    "name": "QFT",
    "rep_kind": "QFT_TEMPLATE",
    "domain_qdt": "reg_phase",
    "codomain_qdt": "reg_phase",
    "params": { "approx_degree": 0, 
                       "do_swaps": true,
                       "inverse": false },
    "cost_hint": { "twoq": 45, "depth": 100 },
    "result_schema": {
      "basis": "Z",
      "datatype": "AS_PHASE",
      "bit_significance": "LSB_0",
      "clbit_order": [
        "reg_phase[0]","reg_phase[1]","reg_phase[2]",
        "reg_phase[3]", "reg_phase[4]","reg_phase[5]",
        "reg_phase[6]","reg_phase[7]","reg_phase[8]",
        "reg_phase[9]"
      ] } }
\end{lstlisting}

An important part of the quantum operator is to provide \texttt{result\_schema}, specifying how a downstream readout should be produced and decoded if measurement occurs. In our example, \texttt{basis = "Z"} selects a computational basis measurement. \texttt{datatype = "AS\_PHASE"} declares that the bitstring encodes a phase value rather than an arbitrary integer. \texttt{bit\_significance = "LSB\_0"} sets the significance so index \(i\) has weight \(2^{i}\). The array \texttt{clbit\_order} gives a sequence of the logical indices whose outcomes are mapped to successive classical bits. Note that the quantum operator descriptor contains no information on gates, pulses, or device details.

\begin{hpcbox}
\textbf{HPC Point-of-View}:  Important examples of operator descriptors in HPC are the PETSc \texttt{Mat} object. It declares the size and mathematical meaning of a linear operator (or shell matrix) while deferring the storage format and kernel to back‑end code. The PETSc solvers ask the descriptor for multiplies and factors. Another example of operator descriptor from HPC is the FFTW plan. In FFTW, a plan is created by providing the library with the desired transform size, dimensionality, placement (in-place/out-of-place), and input/output strides. During planning, FFTW explores many factor-radix decompositions and selects one algorithm schedule. 
\end{hpcbox}

\subsection{Context Descriptors}
\label{sec:context}
A context descriptor is a declarative record that specifies how an operator may be executed, without changing the meaning of the quantum operator. The goal of the context descriptor is to capture execution policies and resources. For instance, the execution policies might include which devices are involved and whether teleportation is allowed for distributed quantum computing, the logical error‑correction code and distance, pulse/control options, or annealer settings. In the middle layer, the objective of the context descriptor is to keep these runtime concerns explicit and portable, so that libraries and backends can consume the same descriptor, bind it to concrete services (communication, QEC, pulse, anneal, ...), and realize the operator accordingly, without hidden global or implicit side effects.

Listing~\ref{lst:context} presents a context descriptor for selecting a state vector simulator as execution engine. Note that this context descriptor specifies how to execute the QFT without altering operator semantics. These fields define compilation and execution policy orthogonally to the quantum data type and operator descriptors. Under \texttt{exec}, \texttt{engine = "gate.aer\_simulator"} selects Qiskit's Aer simulator.
The \texttt{target} block constrains compilation to the gate set \texttt{[sx, rz, cx]} and a linear 10-qubit \texttt{coupling\_map} \texttt{(0-1-2-...-9)}, which forces realistic routing and basis decompositions. Omitting this block yields an ideal all-to-all configuration where all the qubits are connected. The \texttt{options} field passes a transpiler setting \texttt{optimization\_level = 2}. 

\begin{lstlisting}[float=tp,
                   language=json,
                   basicstyle=\ttfamily\footnotesize,
                   numbers=none,
                   caption={Context Descriptor},
                   label={lst:context}]
{
  "$schema": "ctx.schema.json",
  "exec": {
    "engine": "gate.aer_simulator",
    "samples": 4096,
    "seed": 42,
    "target": {
      "basis_gates": ["sx", "rz", "cx"],
      "coupling_map": [[0,1],[1,2],[2,3],
       [3,4],[4,5],[5,6],[6,7],[7,8],[8,9]]
    },
    "options": {
      "optimization_level": 2
    } } }
\end{lstlisting}

\begin{hpcbox}
\textbf{HPC Point-of-View}: HPC middle layers use the concept of context descriptors in different ways. For instance, \texttt{MPI\_Info} hints (e.g., for I/O, or communicating network topology information) provide execution context to optimize the selection of communication engine. \texttt{MPI\_Info} hints steer communication without changing its semantics, e.g, point-to-point, collective or one-sided communication. Another HPC example of execution context is the ADIOS 2 IO/Engine selection and parameters. For instance, the application developer can choose BP4/BP5/SST backend and transports. These ADIOS2 execution contexts specify how data moves in ADIOS2, not what variables mean. An additional example of execution context is the HDF5 Property Lists which allow user to choose drivers, chunking, compression, and MPI‑IO settings.
\end{hpcbox}

\subsubsection{Orthogonal Context Services}
Orthogonal Context Services are system‑level capabilities that are separate from an operator’s mathematical meaning but necessary to run programs on real hardware. In the middle layer, they are passed in explicitly and invoked via service calls, so libraries stay portable and composable while the caller controls resources and platform‑specific behavior.  These service libraries are called \emph{orthogonal} because any algorithmic library can consult them when needed without changing the operator semantics or seizing global state. Examples for this basic service libraries include quantum communication with teleportation and remote operations between devices, error correction with logical encoding/decoding, and “apply logical operator” mapping to fault‑tolerant implementations, pulse/control with optional pulse context and schedule submission for calibrated, device‑specific realizations, and annealing submission to run Ising/QUBO operator descriptors on annealers and return samples/energies.

\subsubsection{Error Correction Codes}
An important role of the middle layer is how to express QEC. In this middle layer, we treat error correction as an execution context. The middle layer records the error policy in a QEC context (for instance, providing code family, distance, decoder choice, frame-tracking, and optional layout hints). At realization time, an orthogonal QEC service binds logical registers (one logical qubit may span dozens of physical qubits under QEC) to patches, inserts syndrome-extraction rounds and fault-tolerant gadgets and chooses a decoder.
The backends are responsible for compiling this fault-tolerant schedule to the target. This separation preserves composability and portability: the same logical program runs unmodified with or without QEC, and across devices, by swapping only the context descriptor.

Listing~\ref{lst:error_ctx_JSON} shows how to include a mechanism for quantum error correction. Beyond the execution policy, the context can carry an explicit \texttt{qec} block that specifies error-correction policy orthogonally to program semantics. Here, \texttt{code\_family = surface} with \texttt{distance = 7} requests a distance-7 surface code, while \texttt{allocator = auto} delegates patch placement and ancilla management to the runtime. The \texttt{logical\_gate\_set} constrains synthesis to fault-tolerant primitives. In this way, quantum operators remain purely logical and operator descriptor remain unchanged. Downstream, the backend consults \texttt{qec} to insert syndrome extraction, fault-tolerance gadgets, decoding, and frame updates.
\begin{lstlisting}[float=tp,
                   language=json,
                   basicstyle=\ttfamily\footnotesize,
                   numbers=none,
                   caption={Error correction policy as specified in the QEC context},
                   label={lst:error_ctx_JSON}]
{
  "$schema": "ctx.schema.json",
  "exec": { ... },
  "qec": {
    "code_family": "surface",
    "distance": 7,
    "allocator": "auto",              
    "logical_gate_set": ["H","S","CNOT","T","MEASURE_Z"],
    ... },
    "extensions": { ... }
  } }
\end{lstlisting}

\subsection{Quantum Algorithmic Libraries}
The algorithmic libraries in the middle layer are reusable collections of logical transformations that act on typed quantum data. They expose these transformations as Quantum Operator Descriptors and remain agnostic to hardware. When needed, they consult orthogonal context services to get information about communication, error correction, pulse/anneal, etc. 

The quantum algorithmic libraries provide commonly used transformations for arithmetic (addition, modular multiplication and exponentiation, comparison), boolean/conditional (controls, predicates, multiplexers, controlled‑Swap), phase/measurement (QFT, controlled‑phase/kickback gadgets, SWAP test, QPE scaffolding, expectation/estimation helpers), and quantum state preparation (Hadamard gates, amplitude encoding, angle encoding).

The quantum algorithmic libraries provide APIs for the construction of quantum operator descriptions, helpers for their composition and inversion, support for late‑binding, and result‑schema helpers for measurements. The quantum algorithmic libraries provide means for validation, including quantum data types compatibility check, and non-interference rules (no hidden measurement/reset). The libraries could also add metadata such as cost hints (e.g. depth, two‑qubit count), and provenance. Finally, realization hooks are provided to rules that lower a quantum operator descriptor to a target‑specific form (gate list, pulse schedule, anneal submission) when the caller supplies a backend/context.

In our work, the algorithmic libraries can be implemented as a Python package that builds and validates JSON artifacts, instead of, for instance, emitting platform-dependent circuits. Each operator is a pure constructor with JSON schema and semantic checks, optional cost-hint estimators, and helpers to attach result schemas. Composition is just a list of descriptors with utilities to check quantum data type compatibility and enforce no hidden measurement/reset. In our implementation, we use a packaging utility to finally combine the quantum data type, operators, and optional context into a submission bundle (\texttt{job.json}).

\section{Proof-of-Concept Implementation}
We implement a proof-of-concept of the middle layer and two backends (IBM Qiskit Aer simulator and D-Wave Ocean) and use the Max-Cut problem to demonstrate the hardware-agnostic properties of our design. For an undirected weighted graph $G=(V, E, w)$, the \emph{Max-Cut} is the partition $V=S\cup \bar S$ of $G$ that maximizes the total weight of edges crossing the cut, i.e., $\max_{S\subseteq V}\sum_{(i,j)\in E} w_{ij}\,\mathbf{1}[i\in S,\,j\in \bar S]$. The problem is NP-hard and appears in clustering and community detection, circuit/layout optimization, and network design. 

We solve the Max-Cut for the 4-node cycle with uniform weights, $w=1$, on two backends that share the same quantum data type \cite{adedoyin2018quantum}. It declares four decision variables with \texttt{encoding\_kind = ISING\_SPIN}, i.e., logical spins $s_i\in\{-1,+1\}$ represented as Boolean readouts. The fields \texttt{id} (\texttt{ising\_vars}) and \texttt{name} (\texttt{s}) identify the logical register; \texttt{width = 4} fixes its size. \texttt{bit\_order = LSB\_0} pins the index-to-significance mapping so bit $i$ is the $i$th character in the decoded string. \texttt{measurement\_semantics = AS\_BOOL} specifies that Z-basis outcomes are decoded as $\{0,1\}$ labels.

The two backends use different operator descriptors for the same typed problem. The gate backend consumes a QAOA descriptor sequence, while the annealer backend consumes a single Ising descriptor. 

\begin{figure}[t]
    \centering
    \includegraphics[width=\linewidth]{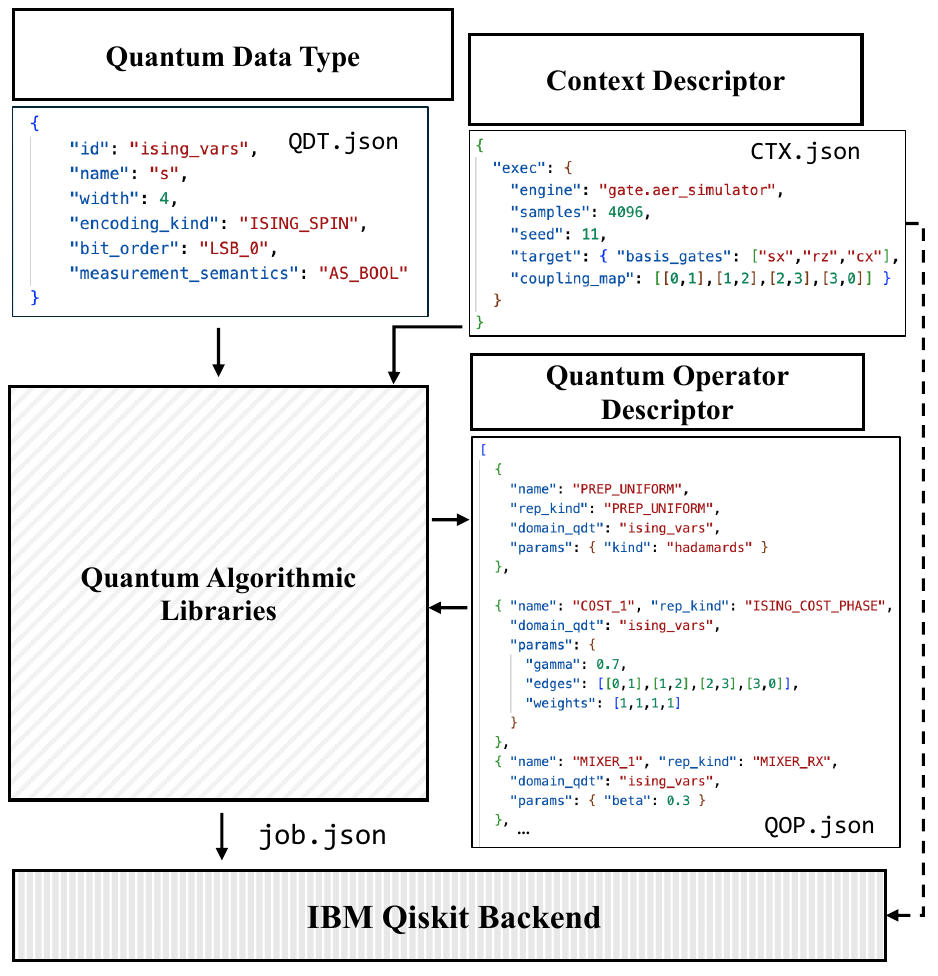}
    \caption{Diagram of the workflow for solving the Max-Cut on IBM Qiskit Aer state vector simulator.}
    \label{fig:qiskit_maxcut}
\end{figure}

We show the QAOA, gate path and Qiskit backend solution in Fig. \ref{fig:qiskit_maxcut}. The operator list for the QAOA comprises different opponent components as prescribed by the algorithm \cite{hidary2021quantum}. \texttt{PREP\_UNIFORM} on \texttt{ising\_vars} requests uniform state preparation (Hadamard gates applied to each qubit). Each \texttt{ISING\_COST\_PHASE} carries a phase angle $\gamma$ and the problem graph (\texttt{edges}, \texttt{weights}). Each \texttt{MIXER\_RX} carries the angle $\beta$. A final \texttt{MEASUREMENT} (omitted in the Fig. ) attaches an explicit \texttt{result\_schema}. 

For executing the Qiskit Aer, the algorithmic library emits a QAOA stack of operator descriptors, shown in the bottom right panel of Fig. \ref{fig:qiskit_maxcut}. It produces an operator for the quantum state preparation, a cost layer parameterized, a mixer layer, and a final measurement. The measurement carries an explicit result schema. Execution is steered by an \texttt{exec} context (engine, samples/shots, seed, optional target constraints). 

The context descriptor for the Qiskit backend specifies the execution policy for the QAOA (gate path) without changing algorithm semantics.  It is similar to the descriptor discussec in Section~\ref{sec:context}, but uses
a four-qubit ring \texttt{coupling\_map} $(0\!\!-\!\!1\!\!-\!\!2\!\!-\!\!3\!\!-\!\!0)$.

\begin{figure}[t]
    \centering
    \includegraphics[width=\linewidth]{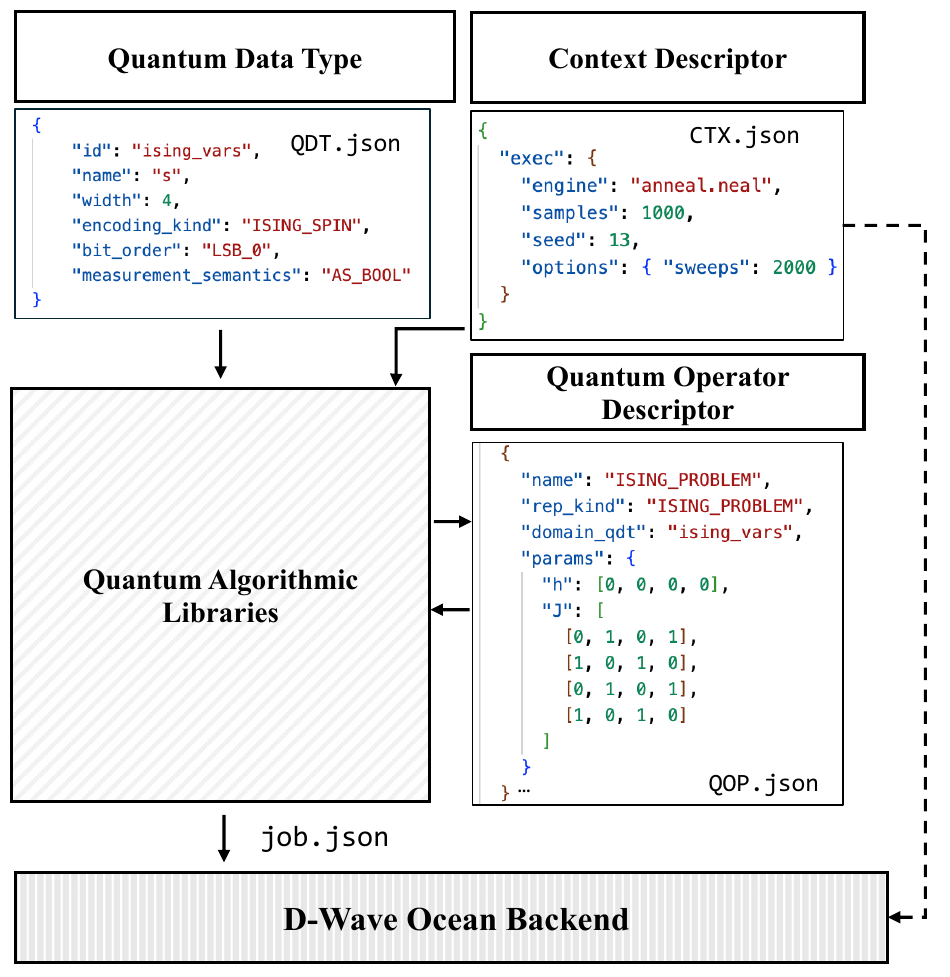}
    \caption{Diagram of the workflow for solving the Max-Cut on D-Wave Ocean Neil simulator.}
    \label{fig:ocean_maxcut}
\end{figure}

We show the quantum annealer path and D-Wave Ocean backend in Fig. \ref{fig:ocean_maxcut}.
The library emits a single \texttt{ISING\_PROBLEM} descriptor (equivalently a QUBO/BQM) specifying $(h,J)$ for the same graph. This operator descriptor declares an Ising problem on the logical register \texttt{ising\_vars}. The \texttt{rep\_kind = ISING\_PROBLEM} indicates that the intent is to define the Ising energy $E(s)=\sum_i h_i s_i + \sum_{i<j} J_{ij} s_i s_j$ over spins $s_i\in\{-1,+1\}$. Here \texttt{h} is the zero vector and \texttt{J} is a symmetric $4\times4$ matrix with unit couplings on edges \((0,1),(1,2),(2,3),(3,0)\).

The  execution context descriptor sets for execution on the D-Wave Ocean simulator, as shown in the top right panel of Fig. \ref{fig:ocean_maxcut}. The block \texttt{"contexts": \{ "anneal": \{ \dots \}\}} specifies \texttt{num\_reads = 1000}, the number of independent samples (anneals) to draw.

When comparing the results of two backend, both runs produce the optimal cut assignments 1010 and 0101 (cut =4). In our basic settings and proof-of-concept implementation, the expected cut (the average cut value over all returned bitstrings, weighted by how often each string was observed) is approximately 3.0-3.2.

\section{Related Work}
Research on high-level abstractions and quantum data models began with early quantum programming languages. For instance, Quipper introduces a higher-order functional Domain Specific Language that distinguishes parameters from inputs and supports hierarchical circuit composition, and quantum data types~\cite{green2013quipper}. ProjectQ generalized this idea with a Python-embedded DSL and modular, backend-agnostic engines~\cite{steiger2018projectq}. Q\# proposes language-level types and functors to support safe classical-quantum composition at a hardware-agnostic middle layer \cite{svore2018q}. Silq raises abstraction by automating uncomputation and type-directed semantics~\cite{bichsel2020silq}. For program and data models, OpenQASM3 formalizes timing, control flow, classical datatypes, and registers for real-time hybrid execution, providing a schema to express circuits, measurements, and payload interface~\cite{cross2022openqasm}.

Several works have focused on the middle layer by providing an IR that separates the quantum application from the hardware details. XACC~\cite{mccaskey2018language} is an example of such frameworks, targeting the integration of HPC-quantum systems. It uses a polymorphic IR and an extensible IR, and the high-level program can be compiled on different backends. Another example is QCOR~\cite{mintz2020qcor}. It builds on XACC to support a higher-level C++ quantum language embedded via library extensions. QCOR supports hardware-agnostic quantum acceleration in classical workflows. t|ket>~\cite{sivarajah2020t} is a quantum optimizing compiler and software development kit that is language-agnostic and retargetable. At its core, a hardware-independent IR for quantum circuits allows the same high-level circuit to be compiled for many devices by simply switching the backend.  

Inspired by classical compiler design, standardized IRs for quantum programs have been proposed to enable portability and interoperability. For example, QIR (Quantum Intermediate Representation) was introduced in 2020-2021 (by Microsoft and the QIR Alliance) as an LLVM-based IR for quantum computing~\cite{cardama2025review}. It embeds quantum instructions into an LLVM IR extension, allowing quantum programs to be written in high-level languages. Similarly, QSSA (Quantum Static Single Assignment) form~\cite{peduri2022qssa} has been investigated to apply compiler optimizations in a static-analysis-friendly way, and quantum MLIR~\cite{mccaskey2021mlir,ittah2022qiro} to extend LLVM’s MLIR to support multi-level quantum code representation.

The integration of HPC and quantum computing tasks into combined frameworks has been examined by
Saurabh, Jah, and Luckow~\cite{saurabh2023a}.  They propose another type of middleware based on
workflow management treating the quantum tasks as self-contained components defined by means
external to the proposed middleware, for example, using Qiskit or QASM.
As illustrated in Fig.~\ref{fig:middle_layer} our quantum middle layer could be used to define
such quantum tasks.
Pilot-Quantum~\cite{mantha2025a} is a workload, resource and task management system built along
the ideas of~\cite{saurabh2023a} using Apache Ray, Dask, and SLURM as components allowing to
execute quantum tasks on remote hardware or local simulators.  It can integrate
circuit-cutting techniques that can dynamically create new smaller quantum tasks.  Similarly in
spirit, Pilot-Quantum could instantiate different suitable quantum circuits by calling our
quantum middle layer to concretize agnostic quantum tasks for different available quantum
hardware or simulators at run-time.
Brown, Meller, and Richings define a C-like API called CQ~\cite{brown2025a} to simplify the
use of quantum computing elements to accelerate HPC applications.  CQ focuses on standardization
of the interface between a powerful host and a classic coprocessor controlling the quantum
hardware allowing an HPC application to call upon quantum computing subroutines at run-time.
CQ does provide API calls to construct both gate-based and analog quantum circuits, but
explicitly cautions the user that concrete vendor- or hardware-specific CQ implementations
are likely to extend the standard API.  As such, while the interface to the coprocessor is
hardware-agnostic, the definitions of the quantum algorithms are not necessarily so.

\section{Discussion and Conclusion}
The quantum software ecosystem has grown considerably in the last decades, leading to formulation and development of instruction sets, compilers, and programming frameworks, and associated abstractions. In this work, we focus on the development of a structured, portable, and composable middle layer. Our quantum middle layer consists of four main components. First, quantum data type descriptors give registers explicit meaning (width, encoding, qubit ordering, measurement interpretation). Second, quantum operator descriptors name logical transformations with parameters and optional cost hints, independent of realization. Third, a context descriptor describes execution policy without changing semantics. Finally, algorithmic libraries consume quantum data types and produce quantum operator descriptor sequences for algorithms. In the algorithmic libraries, a packaging step bundles quantum data type, operators, and optional context for submission to a backend.

Our blueprint with orthogonal execution context is similar to MPI datatypes and communicators, PETSc’s separation of linear operators from solvers, FFTW transform descriptors (the plans), and ADIOS/HDF5 engines and property lists. 

 A small proof-of-concept shows that the same typed problem can be realized on a gate simulator (IBM Qiskit Aer state vector simulator) and an annealer workflow (D-Wave Ocean neal optimizer) by changing only the context and operator formulation. In our proof-of-concept implementation, we use JSON files for quantum data type, operator, and context descriptors. A production path would introduce a surface language whose compiler front end emits a portable middle-layer IR as MLIR dialects that encode our model. The MLIR dialects live in the front/mid-end and carry verifier checks and generic passes, such as parameter binding, cost estimation, and composition. The backend then implements lowerings to target forms, e.g., gate/pulse dialects (or Qiskit/Cirq APIs), BQM/Ising for annealers, and CV/control dialects, and the runtime that submits jobs to specific platforms. 

Another potential implementation is to use Protocol Buffers (\texttt{protobuf}) as the canonical IR by defining messages for quantum data types, quantum operator descriptors, and top-level bundles. The advantages are that \texttt{protobuf} provides strong typing, fast binary serialization, and code generation in Python, C++, and Rust. It also supports \texttt{gRPC} transport for remote backends. 

In this work, we avoided prescribing standards at this stage because the quantum stack is still changing quickly. Devices, control models (gate, pulse, CV, anneal), and algorithmic idioms evolve too fast for a fixed committee interface to remain useful. Instead, we focused on backend-neutral contracts. This lets ecosystems interoperate while leaving room for new hardware features and execution policies to appear without breaking existing code.

\begin{acks}
This research is supported by the European Commission under the Horizon project OpenCUBE (GA-101092984).
\end{acks}

\bibliographystyle{ACM-Reference-Format}
\bibliography{QuantumLibrary}

\end{document}